\documentclass[12pt]{article}
\usepackage{amsmath}
\usepackage{graphicx}
\usepackage{enumerate}
\usepackage{natbib}
\usepackage{url} 
\usepackage[normalem]{ulem}

\usepackage[hidelinks]{hyperref}

\input{definition}
\usepackage{xr}

\makeatletter
\newcommand*{\addFileDependency}[1]{
  \typeout{(#1)}
  \@addtofilelist{#1}
  \IfFileExists{#1}{}{\typeout{No file #1.}}
}
\makeatother

\newcommand{\blind}{1}

\begin{document}

\def\spacingset#1{\renewcommand{\baselinestretch}%
{#1}\small\normalsize} \spacingset{1}

\date{\empty}

\if1\blind
{
  \title{\bf Robust Inference for High-dimensional Linear Models with Heavy-tailed Errors via Partial Gini Covariance}
  \author{Yilin Zhang\\
    Department of Statistics, London School of Economics and Political Science\\
    Songshan Yang \\
    Institute of Statistics and Big Data,
    Renmin University of China \\
    Yunan Wu \\
    Yau Mathematical Science Center,
    Tsinghua University \\
    and \\
    Lan Wang \\
    Department of Management Science,
    Miami Herbert Business School, \\
    University of Miami \\ 
    }
  \maketitle
} \fi

\if0\blind
{
  \bigskip
  \bigskip
  \bigskip
  \begin{center}
    {\LARGE\bf 
    Robust Inference for High-dimensional Linear Models with Heavy-tailed Errors via Partial Gini Covariance}
\end{center}
  \medskip
} \fi

\bigskip
\begin{abstract}
This paper introduces the partial Gini covariance, a novel dependence measure that addresses the challenges of high-dimensional inference with heavy-tailed errors, often encountered in fields like finance, insurance, climate, and biology. Conventional high-dimensional regression inference methods suffer from inaccurate type I errors and reduced power in heavy-tailed contexts, limiting their effectiveness. Our proposed approach leverages the partial Gini covariance to construct a robust statistical inference framework that requires minimal tuning and does not impose restrictive moment conditions on error distributions. Unlike traditional methods, it circumvents the need for estimating the density of random errors, and enhances the computational feasibility and robustness. Extensive simulations demonstrate 
the proposed method’s superior power and robustness over
standard high-dimensional inference approaches,
such as those based on the debiased Lasso.
The asymptotic relative efficiency analysis provides additional theoretical insight on
the improved efficiency of the new approach in the heavy-tailed setting. 
Additionally, the partial Gini covariance extends to the multivariate setting, enabling chi-square testing for a group of coefficients. 
We illustrate the method’s practical application with a real-world data example.
\end{abstract}

\noindent%
\vfill

\newpage
\spacingset{1.75} 
\section{Introduction}
\label{sec:intro}

Modern real-world big data, prevalent in fields such as finance, insurance, climate, and biology, often exhibits 
heavy-tailed behavior \citep{Peinke2004,Ahmad2020,Gupta2022},
characterized by the presence of outliers or extreme values.
When faced with heavy-tailed errors, conventional high-dimensional regression methods frequently yield biased estimates, imprecise inference, and untrustworthy risk assessment. For example, \citet{cavanaugh2015probability} observed the heavy-tailed phenomenon while analyzing the daily precipitation data from more than 22,000 global stations.
Neglecting the contamination of heavy-tailed random errors in such data analysis could result in an underestimation of the likelihood
and severity of the events of interest, thereby significantly impacting disaster preparedness and response strategies.
While extensive research efforts have been dedicated
to statistical inference with heavy-tailed random errors for low-dimensional data
 \citep{Kotchoni2012,Davidson2012,Madeira2015}, 
 studies on high-dimensional inference with heavy-tailed errors remain limited. This is a crucial gap in the literature, as many real-world applications, such as those in finance, insurance, climate, and biology, involve high-dimensional datasets with heavy-tailed behavior.

In this paper, we introduce the {\it partial Gini covariance}, a novel dependence measure
that serves as a foundation for robust statistical inference in high dimensions.
Standard high-dimensional inference procedures 
based on the debiased Lasso \citep{van2014asymptotically, javanmard2014confidence, zhang2014confidence} are observed to suffer from inflated type I errors and substantial power loss with heavy-tailed random errors,
as demonstrated in the Monte Carlo studies in Section~\ref{sec:4}.
Our new inference procedure addresses three
key challenges. First, it effectively handles  heavy-tailed random errors, including Cauchy distribution errors, which are often overlooked in the existing
literature on high-dimensional robust regression.
Second, it is straightforward to implement with minimal tuning requirements, avoiding the complexities involved in selecting the regularization parameters in high-dimensional regression with heavy-tailed errors.
Lastly, it enables valid inference on low-dimensional target regression coefficients
without estimating the random errors' density function.

Recent advances in robust high-dimensional estimation have primarily focused on Huber regression 
\citep{loh2021scale,wang2021new,sun2020adaptive}. 
While these approaches relax the sub-Gaussian error assumption, they still require certain moment conditions. 
For instance, \citet{sun2020adaptive} assumes the existence of the $(1+\delta)$-th moment of the random error. The robust inference procedure developed in \citet{han2022robust} 
requires the existence of the first moment of the random error. These moment conditions exclude many heavy-tailed distributions, 
such as Cauchy or Laplace distributions, that are fundamental to traditional robust statistics.
Moreover, high-dimensional Huber-type regression requires tuning two parameters—one for regularization and one for robustification—making it computationally demanding.

Consider the high-dimensional linear 
regression model 
\begin{equation}
\label{equation:linearmodel}
Y = \x\trans\bbeta^{\ast} +  \varepsilon, 
\end{equation}
where $\x = (X_{1}, \ldots, X_{p})\trans$ is the $p$-dimensional vector of covariates, 
$\bbeta^{\ast} = (\beta_{1}^{\ast} ,\ldots, \beta_{p}^{\ast} )\trans\in \mR^{p}$ is the $p$-dimensional coefficient, $\varepsilon$ denotes the random error independent of $\x$. 
Without loss of generality, we assume $\E(\x) = \vz_{p}$, where $\vz_{p}$ denotes the $p$-dimensional zero vector and the intercept is absorbed into $\varepsilon$. 
The number of covariates $p$ may substantially exceed the sample size $n$,
and notably we do not impose any moment condition on $\varepsilon$.
For an arbitrary $1\leq k\leq p$, 
we aim to test the significance of target parameter $\beta_{k}^{\ast}$ in the presence of the high-dimensional nuisance parameter $\bbeta^{\ast}_{-k}\defby(\beta_{1}^{\ast},\ldots,\beta_{k-1}^{\ast},\beta_{k+1}^{\ast},\ldots,\beta_{p}^{\ast})\trans$, and extend this
to simultaneous testing the significance of a group of coefficients.
Let $\{(\x_{i}, \varepsilon_{i})\}_{i=1}^{n}$ be independent copies of $(\x,  \varepsilon)$. 
The observed data consist of $\{(Y_{i}, \x_{i})\}_{i=1}^{n}$ with $Y_{i} = \x_{i}\trans\bbeta^{\ast} + \varepsilon_{i}$.

The Gini covariance is a useful measure of the dependence between random variables from heavy-tailed distributions, and it is widely applied in areas such as economics (e.g., studying the relationship between income and expenditure) and social sciences \citep{yitzhaki2013gini}.
In this paper, we introduce the concept of partial Gini covariance, partly inspired by the Rank Lasso procedure \citep{wang2020tuning} which incorporates the Gini covariance \citep{schezhtman1987measure} into its loss function for robust estimation. 
We show that the partial Gini covariance, introduced in Section 2, acts as a robust decorrelated score function satisfying Neyman's orthogonality condition, and serves as the foundation for robust statistical inference in high-dimensional regression. 
We propose an estimator for the partial Gini covariance in the high-dimensional setting, and based on the estimation, we construct a hypothesis testing procedure and investigate the limiting distributions of the proposed test statistic under both the null hypothesis and the local alternatives.

Our approach differs from inference methods based on Huber regression in that it does not require moment conditions on random errors and it 
circumvents the complexity of simultaneously
tuning two regularization parameters.
Unlike the debiased Rank Lasso method, which is informally discussed in \cite{fan2020comment} for the low-dimensional scenario, we address the more challenging high-dimensional inference problem. Our proposed statistic converges to a normal distribution, and its implementation does not need to estimate the density of the random error. 
Furthermore, by exploring the double summation expression of the partial Gini covariance, we have considerably relaxed the constraints on the sample size $n$ and the dimension $p$. Existing quantile-based debiased methods implicitly require $p = o\{\exp(n^{1/5})\}$ when the random errors follow a sub-Gaussian distribution \citep{belloni2014uniform, zhao2014general, bradic2017uniform}, whereas our method only requires  $p = o\{\exp(n^{1/3})\}$, despite the discontinuous nature of the sample loss function. 
Our requirement is also less stringent than that for the debiased Huber regression, which requires $p = o\{\exp(n^{\alpha/3 })\}$ for some $\alpha\in(0,1)$ \citep{han2022robust}.
In addition, we provide the asymptotic relative efficiency (ARE) between the proposed method and the existing high-dimensional inference methods including \cite{zhang2014confidence,ning2017general,ma2017variable}.
Our ARE analysis provides additional theoretical insight on the improved efficiency of our approach in the heavy-tailed setting.
As illustrated in Figure \ref{fig:power},
the efficiency improvement can be substantial, with power performance exceeding four times those of the existing methods when the random error follows a Cauchy distribution.


 Additionally, we extend the concept of partial Gini covariance to the multivariate case and establish an asymptotic chi-square test to test the significance of a multivariate coefficient vector. We conduct extensive simulation studies and compare our method with three prominent existing methods, including
the debiased Lasso \citep{van2014asymptotically, javanmard2014confidence, zhang2014confidence}, 
 the partial Pearson covariance-based method \citep{ning2017general} and partial quantile covariance-based methods \citep{li2015quantile,ma2017variable}. Our results demonstrate that our approach exhibits superior robustness and efficiency in handling high-dimensional data with heavy-tailed random errors.

The remainder of the paper is organized as follows.
Section \ref{subsec:2.1} introduces the concept of partial Gini covariance and establish the corresponding estimation method. 
In Section \ref{sec:2}, we  propose a statistical inference procedure for high-dimensional linear regression and establish asymptotic properties of the test statistic under both the null and alternative hypotheses. Furthermore, we extend the notion of partial Gini covariance to the multivariate scenario and construct a chi-square test.
In Section \ref{sec:3}, we compare our methods with standard
approaches for high-dimensional inference and analyze the asymptotic relative efficiency. 
Section \ref{sec:4} provides comprehensive simulation studies 
for testing the significance of both univariate and multivariate coefficients.
We illustrate the application of the new methods using a real-world data example in Section \ref{sec:5}. 
Technical proofs are deferred to the Supplementary Material.

\section{Partial Gini Covariance}\label{subsec:2.1}

To construct a test statistic that effectively handles heavy-tailed errors in the high-dimensional setting, we introduce a novel measure called the partial Gini covariance in Section~\ref{PGC1}. We then discuss how to estimate the partial Gini covariance in high dimensional settings in Section~\ref{PGC2}.

\subsection{Partial Gini Covariance and Neyman Orthogonality}
\label{PGC1}

 To introduce the notion of partial Gini covariance, we first recall that
the Gini covariance between two random variables $Y$ and $X$ is defined as
\beqrs
\Gcov(Y,X) \defby \cov\left\{  F_{Y}(Y),X\right\},
\eeqrs
where $F_Y(y)$ is the cumulative distribution function of $Y$,
and $\cov(\cdot, \cdot)$ denotes the covariance function.
It is the basic building block for 
Gini regression \citep{olkin1992gini}, the loss function of which can  be written as
\beqrs
\Gcov(Y-\bbeta\trans\x,Y-\bbeta\trans\x) =\frac{1}{4}
\E\big\{ \abs{(Y_{i}-\bbeta\trans\x_{i}) - (Y_{j}-\bbeta\trans\x_{j})}\big\},
\eeqrs
where $1\leq i, j \leq n$; $Y_{i}$ and $Y_{j}$ are independent copies of $Y$; $\x_{i}$ and $\x_{j}$ are independent copies of $\x$   (e.g., \citet{lerman1984}, Section 2.1.3 of \citet{yitzhaki2013gini}). Gini regression is
known to be equivalent to the popular rank regression
loss function with the Wilcoxon score function 
\citep{jaeckel1972,Hettmansperger1998},
which enjoys superior robustness and
efficiency properties.

For $1\leq k\leq p$, we denote $\z_{k}\defby \x_{-k}$, where $\x_{-k} = (X_{1},\ldots, X_{k-1}, X_{k+1},\ldots,X_{p})\trans$, for notational simplicity.
To measure the dependence between $Y$ and $X_{k}$ adjusting for the confounder $\z_{k}$ in the heavy-tailed case, we introduce an intuitive new concept, {\it partial Gini covariance}.
We first regress $X_{k}$ on $\z_{k}$ and denote $\bgamma_{k}= \arg\min_{\bgamma}\E\left\{(X_{k} - \bgamma\trans\z_{k})^2\right\}$.
Then, we regress $Y$ on $\z_{k}$ using Gini regression and denote $\btheta_{G,k}=\arg\min_{\btheta} \E\left\{(Y - \btheta\trans\z_{k}) F_{ Y - \btheta\trans\z_{k}}( Y
\right.$\\$\left.
- \btheta\trans\z_{k})\right\}$,  where $F_{ Y - \btheta\trans\z_{k}}(\cdot)$ represents the cumulative distribution function of the random variable $Y - \btheta\trans\z_{k}$. 
The partial Gini covariance between $Y$ and $X_{k}$ given $\z_{k}$ is defined as
\beqr\label{equation:partailgini}
\pGcov(Y, X_{k}\mid \z_{k}) = \cov\left\{ F_{Y  - \btheta_{G,k}\trans\z_{k}}(Y -\btheta_{G,k}\trans\z_{k}), X_{k} - \bgamma_{k}\trans\z_{k}\right\}.
\eeqr 
The partial Gini covariance thus is the Gini covariance between $Y - \btheta_{G,k}\trans\z_{k}$ and $X_{k} - \bgamma_{k}\trans\z_{k}$.
We observe that $\btheta_{G,k}$ satisfies an
important moment condition when $E(\z_{k}) = \vz_{p-1}$ and $\varepsilon_{1}-\varepsilon_{2}$ follows continuous distribution:
\beqr\label{equation:momontzero}
\E\left\{\z_{k} F_{ Y - \btheta_{G,k}\trans\z_{k}}( Y - \btheta_{G,k}\trans\z_{k})\right\} = \vz_{p-1}.
\eeqr
The equation holds since the derivative of $\E\left\{(Y - \btheta\trans\z_{k}) F_{ Y - \btheta\trans\z_{k}}( Y - 
\btheta\trans\z_{k})\right\}$ with respect to $\btheta$ at $\btheta_{G,k}$ equals zero, which is proved in the Supplementary Materials.
As a result, the Gini covariance between $Y - \btheta_{G,k}\trans\z_{k}$ and covariate $\z_{k}$ equals zero.

It is clear that $Y$ has a bounded influence 
for the partial Gini covariance as its effect is through $F(\cdot)$, which is bounded.
Meanwhile, Lemma \ref{lemma:neyman_orthogonality} below
shows the partial Gini covariance in \eqref{equation:partailgini} enjoys a property called Neyman orthogonality, which 
ensures 
small errors in estimating the nuisance parameters do not significantly affect the estimation of the parameter of interest
\citep{Neyman1959, Neyman1979, chernozhukov2015valid}.
The bounded influence and the Neyman orthogonality together provide the foundation for robust inference in high dimensions.

Define $M(\btheta, \bgamma) = \cov\left\{ F_{Y  -  \btheta\trans\z_{k}}(Y -\btheta\trans\z_{k}), X_{k} - \bgamma\trans\z_{k}\right\} $.
{\lemm\label{lemma:neyman_orthogonality} 
 Suppose Condition \ref{condition:continuous_error} in Section \ref{sec:3.1} holds.
When $\beta_{k}^{\ast} = 0$, we have
\beqrs
\frac{\partial M(\btheta, \bgamma)}{\partial \btheta}  \Big\lvert_{(\btheta,\bgamma) = (\btheta_{G,k}, \bgamma_{k})}  = \vz_{p-1} 
~\textrm{ and } ~
\frac{\partial M(\btheta, \bgamma)}{\partial \bgamma}  \Big\lvert_{(\btheta,\bgamma) = (\btheta_{G,k}, \bgamma_{k})}  = \vz_{p-1}.
\eeqrs}

Lemma \ref{lemma:neyman_orthogonality} indicates that the estimation of  $\pGcov(Y, X_{k}\mid \z_{k})$  is locally insensitive to small perturbations of the parameters $\boldsymbol{\theta}_{G,k}$ and $\bgamma_{k}$. 
As the result,
the estimate for the partial Gini covariance can still achieve root-$n$ asymptotic normality even when the estimation bias 
measured in the $L_2$ norm of $\btheta_{G,k}$ and $\bgamma_{k}$ is of order $O\{(n^{-1/2}\log p)^{1/2}\}$ in high dimensions.




\subsection{The Estimation of Partial Gini Covariance in High Dimensions}
\label{PGC2}


We are interested in the high-dimensional scenario where $p$ is much larger than $n$ ($p\gg n$). To overcome the overfitting due to high dimensionality, we use regularized estimates with sparsity assumptions on both $\boldsymbol{\theta}_{G,k}$ and $\boldsymbol{\gamma}_{k}$. \\
\textbf{Estimation of $\btheta_{G,k}$.} 
We adopt the regularized Gini regression (or equivalently rank regression) to estimate $\btheta_{G,k}$ in the high-dimensional setting. Specifically, 
the estimate of $\btheta_{G,k}$ is defined
as 
$\wh\btheta_{G,k}\defby \arg\min_{\btheta}\calL_{1}(Y, \z_{k}; \btheta)$, 
where 
$\btheta=(\theta_1, \ldots, \theta_{p-1})\trans$ and
$\calL_{1}(Y, \z_{k}; \btheta)$
is the regularized Gini loss function
given by
\beqr
\label{ranklasso}
\calL_{1}\left(Y, \z_{k}; \btheta\right) = \{n(n-1)\}^{-1} \sum_{i=1}^{n}\sum_{j\neq i}^{n}|(Y_i-\btheta\trans\z_{k,i} )
-(Y_j-\btheta\trans\z_{k,j} )| + \lambda_{Y,G,k}\sum_{l=1}^{p-1}\abs{\theta_{l}}.
\eeqr

The estimation error bound for  
the high-dimensional regularized rank Lasso
was derived in \citet{wang2020tuning}.
Rank Lasso enjoys two particularly appealing properties. 
First, it possesses desirable robustness 
and efficiency in the presence of heavy-tailed error contamination.
Second, its implementation is free of tuning 
parameter
in the sense that $\lambda_{Y,G,k}$ could be simulated without knowledge of the random error distribution.
Specifically, let $\s_{n} = -2\{n(n-1)\}^{-1}\sum_{i=1}^{n}\z_{k,i}r_{i}$ denote the subgradient, with $\{r_1,\ldots,r_{n}\}$ following the uniform distribution on the permutations of $\{1, 2, \ldots, n\}$. \cite{wang2020tuning} showed that 
for any given $0<\alpha_{0}<1$ and $c>1$,
the choice of the tuning parameter
$\lambda_{Y,G,k}=cF^{-1}_{\|\s_{n}\|_{\infty}}(1 - \alpha_{0})$, where
$F^{-1}_{\|\s_{n}\|_{\infty}}(1 - \alpha_{0})$ denotes $(1 - \alpha_{0})$-quantile of the distribution of $\|\s_{n}\|_{\infty}$,  ensures  
a desirable estimation accuracy with high probability.
It is noted that the distribution of $\s_{n}$ does not depend on the error distribution.

\noindent\textbf{Estimation of $\bgamma_{k}$.} 
The estimate of $\bgamma_{k}$ is defined as $\wh\bgamma_{k}\defby \arg\min_{\bgamma} \calL_{2}\left(X_{k}, \z_{k}; \bgamma\right)$, where
\beqrs
\calL_{2}\left(X_{k}, \z_{k}; \bgamma\right)  \defby (2n)^{-1}\sum_{i=1}^{n} (X_{k,i} - \bgamma\trans\z_{k,i})^2 +
\lambda_{X,k}\sum_{l=1}^{p-1}\abs{\gamma_{l}}.
\eeqrs
with 
 $\lambda_{X,k}$ being the tuning parameter.\\
\textbf{Estimation of the partial Gini covariance.} Denote $\wh\varepsilon_{k,i} \defby Y_{i} - \wh\btheta_{G,k}\trans\z_{k,i}$ and let $R(\wh\varepsilon_{k,i} )$ be the rank of $\wh\varepsilon_{k,i}$ among $\{\wh\varepsilon_{k,i}, i=1,\ldots,n\}$.
Our estimate for $\pGcov(Y, X_{k}\mid \z_{k})$ is  
\beqr\label{equation:giniestimate}
\wh\pGcov( Y, X_{k}\mid \z_{k}) &\defby& n^{-1}\sum_{i=1}^{n}(X_{k,i} - \wh\bgamma_{k}\trans\z_{k,i})\{R(\wh\varepsilon_{k,i} )/n - 1/2\}  \\
&=& n^{-2}\sum_{i=1}^{n}\sum_{j=1}^{n}\left(X_{k,i} - \wh{\bgamma}\trans\z_{k,i}\right) \left\{
1\left(\wh\varepsilon_{k,i} \geq 
\wh\varepsilon_{k,j} \right) - 1/2\right\}. \nonumber
\eeqr
The penalized Gini regression method used to estimate $\btheta_{G,k}$ and the indicator function in $\wh\pGcov(Y, X_{k}\mid \z_{k})$ both contribute to the robustness of 
the estimated partial Gini covariance.
Furthermore, the estimate is computationally efficient.


Denote $s_{X} \defby \max_{1\leq k\leq p} \|\bgamma_{k}\|_{0}$ and $s_{Y,G} \defby  \max_{1\leq k\leq p} \|\btheta_{G,k}\|_{0}$.
When the covariates follow the sub-Gaussian distribution, 
$\|\wh\btheta_{G,k} - \btheta_{G,k}\|_{2} = O_p(\lambda_{Y,G,k}s_{Y,G}^{1/2})$ and $\lambda_{Y,G,k} =O\{(n^{-1}\log p)^{1/2} \}$ under the lower restricted eigenvalue condition and the random error condition \citep{wang2020tuning}.
Under the lower restricted eigenvalue condition,  $\|\bgamma_{k} - \bgamma_{k}\|_{1} = O_p(\lambda_{X,k}s_{X})$, where $\lambda_{X,k}=O\{(n^{-1}\log p)^{1/2} \}$ \citep{fan2020statistical}.
We relegate the lower restricted eigenvalue condition and the random error condition to the Supplementary Material. 

\section{The New Inference Procedures}\label{sec:2}
\subsection{Testing the Significance of a Target Covariate}
\label{sec:3.1}
We consider the target parameter of interest $\beta_{k}^{\ast}$, with $\z_{k}$ representing the vector of high-dimensional confounders.
Our main objective is to test the following hypothesis:
\begin{equation}
	\label{test}
	H_{0,k}: \beta_k^{\ast} = 0\quad \textrm{v.s} \quad H_{1,k}:  \beta_k^{\ast} \neq 0.
\end{equation}

We impose the following regularity conditions to facilitate our technical derivation.

\begin{enumerate}[label=(C\arabic*$^{}$)]
	\item \label{condition:continuous_error}  Let $F_{\varepsilon_{1} - \varepsilon_{2}}(\cdot)$ denote the distribution function of $\varepsilon_{1} - \varepsilon_{2}$. $F_{\varepsilon_{1} - \varepsilon_{2}}(\cdot)$ is continuous on $\mR$.
\end{enumerate}
\begin{enumerate}[label=(C1$'$)]
\item \label{condition:error_distribution} $F_{\varepsilon}(\cdot)$ and $F_{\varepsilon_{1} -\varepsilon_{2}  }(\cdot)$ are second order differentiable, with first and second order derivatives uniformly bounded by a constant $B_{\varepsilon}>0$. 
\end{enumerate}

\begin{enumerate}[label=(C\arabic*$^{}$)]
	\setcounter{enumi}{1}
	\item \label{condition:sub_gaussion}  There exists some constant $c_{1}>0$ such that $\E\{\exp(t\v\trans\x)\}\leq \exp(c_{1}t^2/2)$ for each fixed $\v\in \calS^{d-1}$ and all $t \in \mR$,
 where
	$\calS^{d-1}$ is the surface of $d$-dimensional unit ball.  
	\item \label{condition:order}
	$s_{X}n^{-1/2}\log p = o(1)$ and $ n^{-1/2}s_{Y,G} \log(p) \{\log(n\vee p)\}^{1/2}  = o(1)$.
\end{enumerate}

Condition \ref{condition:continuous_error} and its strengthen version
 Condition \ref{condition:error_distribution} allow the random error $\epsilon$ to follow a heavy-tailed distribution such as the Cauchy
 distribution, relaxing the standard normality or 
 sub-gaussianity assumption on $\epsilon$. Condition \ref{condition:sub_gaussion} states that the covariates follow the sub-Gaussian distribution, which is common in high-dimensional inference literature \citep{javanmard2014confidence,ning2017general}.
{\color{black}
The sparsity assumption in 
Condition \ref{condition:order} is typically imposed in the inference of high-dimensional quantile regression  \citep{bradic2017uniform} or composite quantile regression \citep{zhao2014general} for controlling the supremum of an empirical process. 
}
Condition \ref{condition:order} also provides relations among the sample size, the dimension of the data, and the sparsity levels of the regression coefficients. When $s_{X}$ and $s_{G,Y}$ are finite, this condition is equivalent to $\log(p) = o(n^{1/3})$. Compared with the conditions for the inference of high-dimensional quantile regression models \citep{zhao2014general,belloni2014uniform}, which require $\log(p) = o(n^{1/5})$ under the sub-Gaussian setting, the condition in our context is notably less restrictive. This relaxation is facilitated by the double summation form in \eqref{equation:giniestimate}, see the proof of Theorem \ref{theorem:normality}.

We first demonstrate that the testing problem in \eqref{test} is equivalent to the following hypothesis testing problem based on the partial Gini covariance:
\beqr
\label{pctest}
H_{0,k}^{G}: \pGcov(Y, X_{k}\mid \z_{k}) = 0\quad \textrm{v.s} \quad H_{1,k}^{G}:  \pGcov(Y, X_{k}\mid \z_{k}) \neq 0.
\eeqr
	To see the above equivalence intuitively, it is known that
 the population parameter
 $\bbeta^{\ast}$ can be expressed as
 $\arg\min_{\bbeta\in\mR^{p}}\E\{ \abs{(Y_{i}-\bbeta\trans\x_{i}) - (Y_{j}-\bbeta\trans\x_{j})}\}$
(see, e.g., \citet{hettmansperger2010robust}).
Hence, we also have
$\bbeta^{\ast}=\arg\min_{\bbeta\in\mR^{p}} \Gcov(Y-\bbeta\trans\x,Y-\bbeta\trans\x)$.
	If $\beta_{k}^{\ast} = 0$, the derivative of $\Gcov(Y - \boldsymbol{\theta}^\top\z_{k}- bX_{k}, Y - \boldsymbol{\theta}^\top\z_{k} - bX_{k})$ with respect to $b$ at $b=0$ equals zero. 
	Following the similar derivation as that in Section S.1 of the Supplementary Material, we can show that
	\beqrs
	\E \left\{\frac{\partial\Gcov(Y -  \btheta\trans\z_{k}- bX_{k}, Y - \btheta\trans\z_{k} - bX_{k}) }{\partial b} \Big\lvert_{(\btheta,b) = (\btheta_{G,k}, 0)} \right\} = 
	-4 \cdot \pGcov(Y, X_{k}\mid \z_{k}).
	\eeqrs
The equivalence of the two hypotheses thus follows by noting that $\pGcov(Y, X_{k}\mid \z_{k})=0 $ when $\beta_{k}^{\ast}=0$.
The the equivalence between \eqref{test} and \eqref{pctest} 
is formally stated in Lemma \ref{lemma:equalb} below.


{\lemm\label{lemma:equalb}  
	Under Condition \ref{condition:continuous_error},
	$\beta_{k}^{\ast} = 0$ if and only if $\pGcov(Y, X_{k}\mid \z_{k})=0$.
}

%


Let $f_{\varepsilon_1-\varepsilon_2}(\cdot)$ be the density function of $\varepsilon_{1} - \varepsilon_{2}$. 
Define $\sigma_{G,k}^2 \defby 12^{-1}\E\left\{\left(X_{ki} - \bgamma_{k}\trans\z_{ki}\right)^2\right\}$.
Theorem~\ref{theorem:normality} below states the asymptotic normality of $\pGcov(Y, X_{k}\mid \z_{k})$ under both the null and local alternative hypotheses.

{\theo{\label{theorem:normality}} Suppose Conditions 
 \ref{condition:error_distribution}, \ref{condition:sub_gaussion} and \ref{condition:order} hold. We also assume that
\beqr\label{equation:se} 
\|\wh\btheta_{G,k}\|_{0}\leq c_{2}s_{Y,G}
\eeqr
and $\|\bgamma_{k}\|_{2}$ is bounded. 
	\begin{enumerate}
		\item[(i)] 	 Under $H_{0,k}$ in \eqref{test}, $n^{1/2}\wh\pGcov(Y, X_{k}\mid \z_{k})\stackrel{d}{\longrightarrow}{\cal N}(0, \sigma_{G,k}^2) $ as $n\to\infty$, where $``\stackrel{d}{\longrightarrow}"$ stands for ``converges in distribution".
		\item[(ii)] Under $H_{1,k, n}: \beta_{k}^{\ast}= n^{-1/2}\beta_{k,0} \textrm{ with } \beta_{k,0} \neq 0$, 
        $n^{1/2}\wh\pGcov(Y, X_{k}\mid \z_{k}) 
        \stackrel{d}{\longrightarrow} \calN( 
		f_{\varepsilon_{1} - \varepsilon_{2}}(0)\beta_{k,0} \\
        \E \left\{ (X_{k}-\bgamma_{k}\trans\z_{k})^2 \right\} , \sigma_{G,k}^2)$, as $n\to\infty$.
\end{enumerate}}



\noindent\textit{Remark.}
Assumption \eqref{equation:se} is commonly adopted for high-dimensional statistical inference \citep{guo2024model,zhao2014general}.
For rank Lasso, we verify that \eqref{equation:se} holds with high probability in the Supplementary Material.

It follows from Theorem \ref{theorem:normality} that the asymptotic variance of partial Gini covariance is $\sigma_{G,k}^2=12^{-1}\E\left\{\left(X_{ki} - \bgamma_{k}\trans\z_{ki}\right)^2\right\}$ under $H_{0,k}$. 
This result has two important implications.
First, the asymptotic variance $\sigma_{G,k}^2$ is independent of the distribution of $Y$, allowing for $Y$ with a heavy-tailed distribution.
Thus it ensures the robustness of the test based on the partial Gini covariance.
Second, the result eliminates the need for error density estimation, distinct from the robust test statistics based on debiased rank Lasso estimator  \citep{fan2020comment}, where the asymptotic variance depends on the density of the random error.

To conduct the hypothesis test \eqref{test},  we employ $\wh\sigma_{G,k}^2 \defby (12n)^{-1}\sum_{i=1}^{n}\left(X_{ki} - \wh\bgamma_{k}\trans\z_{ki}\right)^2$ as the estimate of $\sigma_{G,k}^2$.  
By Corollary 4.3 of \citet{ning2017general}, $\wh\sigma_{G,k}^2$ is a consistent estimate of $\sigma_{G,k}^2$.
We reject $H_{0,k}$ in \eqref{test} at a significance level of $\alpha$ if
\beqrs
n^{1/2}\left|\wh\pGcov(Y, X_{k})/\wh\sigma_{G,k}\right| \geq Z_{\alpha/2},
\eeqrs
where $Z_{\alpha/2}$ is the the upper $\alpha/2$-quantile  of the standard normal distribution. 
The proof of Theorem \ref{theorem:normality} is challenging due to the high dimensionality and the discontinuity of the indicator function. To overcome this obstacle, we employ decoupling techniques for the U process, which effectively manages the discontinuity term. Further details can be found in the Supplementary Material.



\subsection{Simultaneous Testing the Significance of a Group of Covariates}\label{subsec:3.4}
Let $\mathcal{S} = \{k_{1},\ldots, k_{d}\}$ denote the index set of the parameters of interest, where $d\defby\abs{\mathcal{S}}$ denotes the cardinality of the set $\mathcal{S}$.
Let $\boldsymbol{\beta}_{\mathcal{S}}^{\ast} = (\beta_{k_{1}}^{\ast},\ldots, \beta_{k_{d}}^{\ast})\trans$ 
represent the corresponding sub-vector of $\boldsymbol{\beta}^{\ast}$. We now consider testing 
if these coefficients are simultaneously zero:
\beqr\label{equation:multihypo}
H_{0,\calS}: \bbeta_{\calS}^{\ast}= \vz_{d},\quad \textrm{v.s} \quad H_{1,\calS}:   \bbeta_{\calS}^{\ast}\neq \vz_{d}.
\eeqr
To construct the test statistic, we define the multivariate partial Gini covariance by 
\beqrs
\pGcov(Y, \x_{\calS}\mid\z_{\calS}) \defby \cov\left\{ F_{Y  - \btheta_{G,\calS}\trans\z_{\calS}}(Y -\btheta_{G, \calS}\trans\z_{\calS}), \x_{\calS} - \bGam_{\calS}\trans\z_{\calS}\right\},
\eeqrs
where $\x_{\calS} = (X_{k_{1}}, \ldots, X_{k_{d}})\trans \in\mR^{d}$ 
and $\z_{\calS}\in\mR^{p-d}$ is the sub-vector removing $\x_{\calS}$ from $\x$.
Let $\btheta_{G,\calS}\defby \arg\min\limits_{\btheta\in\mR^{p-d}} \E
\left\{(Y - \btheta\trans\z_{\calS}) F_{ Y - \btheta\trans\z_{\calS}}( Y - \btheta\trans\z_{\calS})\right\}$ and $\bGam_{\calS}\defby  \arg\min\limits_{\bGam\in\mR^{(p-d)\times d}}\E\big(\|\x_{\calS} - \bGam\trans\z_{\calS})\|^2_{2}\big)$.
Lemma~\ref{lemma:equal_multi} below is an extension of Lemma~\ref{lemma:equalb}.


{\lemm\label{lemma:equal_multi}  
	Under Condition \ref{condition:continuous_error},
	$\beta_{\calS}^{\ast} = \vz_{d}$ if and only if $\pGcov(Y, \x_{\calS}\mid \z_{\calS})=\vz_{d}$.}

Based on Lemma \ref{lemma:equal_multi}, testing \eqref{equation:multihypo}  is equivalent to testing
\beqrs
H_{0,\calS}^{G}: \pGcov(Y, \x_{\calS}\mid \z_{\calS})= \vz_{d}, \quad \textrm{v.s} \quad H_{1,\calS}^{G}:  \pGcov(Y, \x_{\calS}\mid \z_{\calS})\neq \vz_{d}.
\eeqrs

To construct the test statistics, 
we first apply Gini regression with Lasso penalty and obtain an estimate of $\btheta_{G,\calS}$ as
$\wh\btheta_{G,\calS}\defby \arg\min_{\btheta}\calL_{1}(Y, \z_{\calS}; \btheta)$ 
with the tuning parameter $\lambda_{Y,G,\calS}$.
To estimate $\bGam_{\calS}$, we use node-wise least square regression with Lasso penalty \citep{van2014asymptotically, javanmard2014confidence, zhang2014confidence}.
Let $\wh\bGam_{\calS}\defby (\wh\bgamma_{k_{1}}, \ldots, \wh\bgamma_{k_{d}})$ be the estimate of $\bGam_{\calS}$. For $j=1,\ldots,d$, by using Lasso penalty with the tuning parameter $\lambda_{X,k_{j}}$, 
\beqrs
\wh\bgamma_{k_{j}}\defby \arg\min_{\bgamma} (2n)^{-1}\sum_{i=1}^{n} \|X_{k_{j},i} - \bgamma\trans\z_{\calS,i}\|_{2}^2 + \lambda_{X,k_{j}}\|\bgamma\|_{1}.
\eeqrs

Similarly to the univariate case, the error bounds for $\btheta_{G,\calS}$ and $\bGam_{\calS}$ have been derived in the previous work.
Based on \cite{wang2020tuning}, we have $\|\wh\btheta_{G,\calS} - \btheta_{G,\calS}\|_{2} = O_p(\lambda_{Y,G,\calS}s_{Y,G}^{1/2})$, where $\lambda_{Y,G,\calS} =O\{(n^{-1}\log p)^{1/2} \}$.
According to \cite{fan2020statistical}, we have $\|\wh\bGam_{\calS} - \bGam_{\calS}\|_{1} = O_p(\lambda_{X,\calS} s_{X})$, where $\lambda_{X,\calS}\defby  \max\{\lambda_{X, k_{j}}, j=1,\ldots, d\}= O\{(n^{-1}\log p)^{1/2} \}$.

Define $\varepsilon_{\calS, i} \defby Y_{i} - \wh{\btheta}_{G,\calS}\trans \z_{\calS,i}$. 
The estimate for $\pGcov(Y, X_{\calS}\mid\z_{\calS})$ is  
\beqrs
\wh\pGcov(Y, \x_{\calS}\mid\z_{\calS}) \defby n^{-2}\sum_{i=1}^{n}\sum_{j=1}^{n}\left(\x_{\calS,i} - \wh\bGam_{\calS}\trans\z_{\calS,i}\right) \left\{
1\left(\wh\varepsilon_{\calS,i} \geq 
\wh\varepsilon_{\calS,j} \right) - 1/2\right\}.
\eeqrs
To test (\ref{equation:multihypo}), we propose the following chi-square test statistic
\beqrs
\wh{W}_{G,\calS}\defby\wh\pGcov(Y, \x_{\calS}\mid\z_{\calS})\trans\wh\bSig_{G,\calS}^{-1}\wh\pGcov(Y, \x_{\calS}\mid\z_{\calS}),
\eeqrs
where 
\beqrs
\wh\bSig_{G,\calS}\defby (12n)^{-1} \sum_{i=1}^{n}\left(\x_{\calS,i} - \wh\bGam_{\calS}\trans\z_{\calS,i}\right) \left(\x_{\calS,i} - \wh\bGam_{\calS}\trans\z_{\calS,i}\right)\trans
\eeqrs
is the estimate of the covariance matrix $\bSig_{G,\calS}\defby 12^{-1} \text{cov}\left(\x_{\calS} - \bGam_{\calS}\trans\z_{\calS}\right)$.
The asymptotic distributions for $\wh{W}_{G,\calS}$ under both the null hypothesis and the local alternatives are given in Theorem \ref{theorem:chi_square} below.
{\theo{\label{theorem:chi_square}} Suppose Conditions 
\ref{condition:error_distribution},
\ref{condition:sub_gaussion}, and
\ref{condition:order} are  valid, $\|\wh\btheta_{G,\calS}\|_{0}\leq c_{2}s_{Y,G}$, and  $\|\bGam_{\calS}\|_{F}$ is bounded.
	\begin{enumerate}
		\item[(i)] 	 Under $H_{0,\calS}$, $n\wh{W}_{G,\calS}\stackrel{d}{\longrightarrow}\chi_{d}^{2} $, as $n\to\infty$, where $\chi_{d}^{2}$ stands for the chi-square distribution with $d$ degrees of freedom;
		\item[(ii)] Under  $H_{1,\calS, n}: \bbeta_{\calS}^{\ast}= n^{-1/2}\bbeta_{\calS,0} \textrm{ with } \bbeta_{\calS,0} \neq \vz_{d}$, as $n\to\infty$,
        \beqrs
        n\wh{W}_{G,\calS}\stackrel{d}{\longrightarrow}\chi_{d}^{2}\big\{ 144f_{\varepsilon_{1} - \varepsilon_{2}}(0)^2 \bbeta_{\calS,0}\trans\bSig_{G,\calS}\bbeta_{\calS,0}\big\},
        \eeqrs
where $\chi_{d}^{2}(\mu)$ is the noncentral chi-square distribution with $d$ degrees of freedom and non-centrality parameter $\mu$.
\end{enumerate}}

Theorem \ref{theorem:chi_square} implies that $\wh{W}_{G,\calS}$ has nontrivial power in detecting the local alternatives even if  $\bbeta_{\calS}^{\ast}$ converges to $0$ at the rate of $n^{-1/2}$.
When $\bbeta_{\calS}^{\ast}$ is a nonzero vector of constants (or fixed alternative), the power of the above test converges to one as $n$ diverges.

\section{Power Advantages for Heavy-tailed Error Distributions: Comparison with Existing Methods}\label{sec:3}

In this section, we conduct a detailed power comparison between our proposed testing method and those utilizing partial Pearson covariance and partial quantile covariance. Our analysis reveals that the proposed method can be substantially more efficient than the partial Pearson covariance approach, particularly in scenarios with heavy-tailed random errors. Furthermore, the analysis suggests greater power than the method based on partial quantile covariance under local alternatives.

\subsection{Comparison with the Method Based on Partial Pearson Covariance }\label{subsec:3.1}

We derive the efficiency gain of the new test based on the partial Gini covariance versus the one based on partial Pearson covariance in the presence of heavy-tailed random errors. 
The partial Pearson covariance, proposed by \cite{yule1897theory}, is a classical tool for testing the significance of the coefficient in the linear model.
In the high-dimensional settings, the estimate for partial Pearson covariance is given by
\beqr\label{equation:partialpearson}
\wh\pPcov(Y, X_{k}\mid \z_{k}) = n^{-1}\sum_{i=1}^{n}(Y_{i}-\wh\btheta_{P,k}\trans\z_{ki})(X_{ki}-\wh\bgamma_{k}\trans\z_{ki}),
\eeqr
where $\wh\btheta_{P,k} = \arg\min_{\btheta}(2n)^{-1}\sum_{i=1}^{n} (Y_{i} - \btheta\trans\z_{k,i})^2 + \lambda_{Y,P,k}\|\btheta\|_{1}$ with $\lambda_{Y,P,k}$ being the tuning parameter. 
Under the null hypothesis, \eqref{equation:partialpearson} is asymptotically equivalent to the test statistic proposed by \cite{zhang2014confidence}.
Therefore, the following comparison also includes some earlier testing methods including \cite{zhang2014confidence,van2014asymptotically,javanmard2014confidence}.
Meanwhile, \eqref{equation:partialpearson} is also equivalent to the decorrelated score test of \cite{ning2017general} for the linear regression model.

They further proved that $n^{1/2}\wh\pPcov(Y, X_{k}\mid \z_{k})$ converges to the normal distribution with mean zero and variance $\sigma_{P,k}^2\defby \E(\varepsilon^2)\cdot \E\big\{(X_{k} -\bgamma_{k}\trans\z_{k})^2\big\}$ under $H_{0,k}$, and
proposed to test \eqref{test} based on the 
statistic $n^{1/2}\wh\pPcov(Y, X_{k}\mid \z_{k})/\wh\sigma_{P,k}$,
where $\wh\sigma_{P,k}^2\defby \big\{n^{-1}\sum_{i=1}^{n}(Y_{i}-\wh\btheta_{P,k}\trans\z_{ki})^2\big\}\big\{n^{-1}\sum_{i=1}^{n}(X_{ki}-\wh\bgamma_{k}\trans\z_{ki})^2\big\}$. 
It is important to note that $n^{1/2}\wh\pPcov(Y, X_{k}\mid \z_{k})/\wh\sigma_{P,k}$ is only applicable to test \eqref{test} when $E(\epsilon^2)$ exists. Thus, $\wh\pPcov(Y, X_{k}\mid \z_{k})$ may not perform well when $\epsilon$ is heavy-tailed.

To compare the performance of the tests using $\wh\pPcov(Y, X_{k}\mid \z_{k})$ and $\wh\pGcov(Y, X_{k}\mid \z_{k})$ for \eqref{test}, we derive Pitman's asymptotic relative efficiency (ARE) \citep{pitman1948lecture, noether1955theorem} of these two tests. Pitman's ARE is a useful measure for comparing the performance of two statistical tests in large samples.
It is defined as the ratio of the sample sizes needed by two tests to achieve the same power under a sequence of
local alternatives.

Specifically, we consider model \eqref{equation:linearmodel}. 
The Pitman sequence of the local alternatives is given by:
$$ H_{1,k,n} : \beta_{k}^{\ast} = n^{-1/2}\beta_{k,0},$$
where $\beta_{k,0} > 0$ is a constant. For ease of presentation, we assume that $\z_{k}$ and $X_{k}$ are independent of each other. 
Under this condition,  we can derive Pitman's ARE as follows:
\beqr\label{equation:are}
&&\ARE\Big(\wh\pGcov(Y, X_{k}\mid \z_{k}),
\wh\pPcov(Y, X_{k}\mid \z_{k})
\Big) = 12\var(\epsilon)f_{\varepsilon_{1} - \varepsilon_{2}}^2(0).
\eeqr
The proof is relegated in Supplementary Material.
\eqref{equation:are} indicates that the ARE solely relies on the distribution of $\epsilon$. In Table \ref{table:are}, we present the values of ARE with different distributions of $\epsilon$. It can be observed that $\wh\pGcov(Y, X_{k}\mid \z_{k})$ demonstrates comparable performance with $\wh\pPcov(Y, X_{k}\mid \z_{k})$ when $\epsilon$ follows the normal distribution while exhibits significantly higher power when $\epsilon$ is heavy-tailed.
\begin{table}[htpb!]
	\captionsetup{font = footnotesize}
	\caption{
 {
 The asymptotic relative efficiency for  $\wh\pGcov(Y, X_{k}\mid \z_{k})$ and $\wh\pPcov(Y, X_{k}\mid \z_{k})$ under different distributions of $\varepsilon$.
		We consider 5 different error distributions including, Normal(0,1):
		standard normal distribution; 
	Uniform[0,1]:	uniform distribution on $[0,1]$;
	T$_{3}$(0,1): $t$-distribution with 3 degrees of freedom;
	Exp(1)"	exponential distribution with rate $1$;
 and LogNormal(0,1):
log-normal distribution.
  }
	}
	\label{table:are}  
	\centering
	\scalebox{1}{
	\begin{tabular}{lllllll}
		\hline
		$\varepsilon$ & Normal(0,1) &   Uniform[0,1] & T$_{3}$(0,1) &  Exp(1)&   LogNormal(0,1)\\ 
		\hline
		ARE  & 0.955 & 1.000 &  1.901 & 3.000 & 7.353 \\
		\hline
	\end{tabular}}
\end{table}

\subsection{Comparison with the Method Based on Partial Quantile Covariance}\label{subsec:3.2}

In the context of robust regression, quantile-based methods are among the most common approaches.  Several studies have investigated variable selection within these methods  \citep{wu2009variable,wang2012quantile,lee2014model}. Notably, significant advancements have been made by \citet{li2015quantile} and \citet{ma2017variable}, who introduced the concept of partial quantile covariance to quantify the interplay between variables $Y$ and $X_{k}$, while taking into account the confounding influence of covariates $\mathbf{z}_{k}\in\mathbb{R}^{p-1}$.
The partial quantile covariance is defined as follows:
\beqrs
\pQcov_{\tau}(Y, X_{k}\mid \z_{k}) = \cov\left\{  \psi_{\tau}(Y - \btheta_{Q,k}\trans\z_{k}   -  Q_{\tau, Y-\btheta_{Q,k}\trans\z_{k}}), X_{k} - \bgamma_{k}\trans\z_{k}\right\}, 
\eeqrs
where
$\btheta_{Q,k}\defby\arg\min_{(\btheta,a)}\E\{\rho_{\tau}(Y - \btheta\trans\z_{k} - Q_{\tau, Y-\btheta\trans\z_{k}})\}$.
The partial quantile covariance and partial Gini covariance are closely related. To bridge these two measures, we introduce an intermediate concept. Inspired by \citet{zou2008composite}, who proposed composite quantile regression to simultaneously consider $T$ levels of quantile $\tau_{t}\in[0,1]$ for $t=1,\ldots,T$ simultaneously,  
we define the partial composite quantile covariance as
\beqrs
\pCQcov(Y, X_{k}\mid \z_{k}) \defby \cov\left\{  T^{-1}\sum_{t=1}^{T}\psi_{\tau_{t}}(Y - \btheta_{CQ,k}\trans\z_{k}   -  Q_{\tau_{t}, Y-\btheta_{CQ,k}\trans\z_{k}}), X_{k}- \bgamma_{k}\trans\z_{k}\right\},
\eeqrs
where
$\btheta_{CQ,k}\defby\arg\min_{\btheta}\E\{T^{-1}\sum_{t=1}^{T}\rho_{\tau_{t}}(Y - \btheta\trans\z - Q_{\tau_{t}, Y-\btheta\trans\z_{k}})\}$. In the case where $\tau_{t} = t/(T+1)$, we can establish Proposition \ref{proposition:relationship}.

{\prop{\label{proposition:relationship}}  $\pCQcov(Y, X_{k}\mid \z_{k})\to\pGcov(Y, X_{k}\mid \z_{k})$ as $T\to\infty$. 
}

In this study, we aim to formulate the asymptotic normality of the partial quantile covariance in the high dimensional case, extending the work of \cite{li2015quantile} which was limited to the case where \( p \) is finite.
Denote  
\beqr
\label{pqc}
(\wh{Q}_{\tau, Y - \btheta_{Q, k}\trans\z_{k} } ,\wh\btheta_{Q, k})\defby \arg\min_{\eta,\btheta} n^{-1} \sum_{i=1}^{n} \rho_{\tau}
(Y_{i} - \btheta\trans\z_{ki} - \eta) + \lambda_{Y,Q,k}\sum_{l=1}^{p-1}\abs{\theta_{l}},
\eeqr
where $\lambda_{Y,Q,k}$ is the tuning parameter.
Then we define the estimate for partial quantile covariance in high-dimensional settings is
\beqrs
\wh\pQcov_{\tau}(Y, X_{k}\mid \z_{k}) = n^{-1}\sum_{i=1}^{n}\psi_{\tau}(Y_{i}-\z_{ki}\trans\wh\btheta_{Q,k} - \wh{Q}_{\tau, Y - \btheta_{Q, k}\trans\z_{k}} ) (X_{ki}-\wh\bgamma_{k}\trans\z_{ki}).
\eeqrs
Let $s_{Y,Q} \defby  \max_{1\leq k\leq p} \|\btheta_{Q,k}\|_{0}$
  be the sparsity level for quantile regression. We impose the following conditions to establish the asymptotic normality of $\wh\pQcov_{\tau}(Y, X_{k}\mid \z_{k})$:
\begin{enumerate}[label=(C\arabic*$^*$)]
	\setcounter{enumi}{2}
	\item \label{condition:error_bound_quantile} $\|\wh\bgamma_{k} - \bgamma_{k}\|_{1} = O_p(\lambda_{X,k} s_{X})$,
	$\|\wh\btheta_{Q,k} - \btheta_{Q,k}\|_{2} + 
	\|\wh{Q}_{\tau, Y - \btheta_{Q, k}\trans\z_{k}} - Q_{\tau, Y - \btheta_{Q, k}\trans\z_{k}}\|_{2} = O_p(\lambda_{Y,Q,k} s_{Y,Q}^{1/2})$, 
	$\|\wh\btheta_{Q,k}\|_{0}\leq c_{3}s_{Y,Q}$. 
	$\lambda_{X,k}=O\{(n^{-1}\log p)^{1/2} \}$ and $\lambda_{Y,Q,k} =O\{(n^{-1}\log p)^{1/2} \}$.
	\item \label{condition:order_quantile} 
	$ s_{X}n^{-1/2}\log p = o(1)$ and
	$s_{Y,Q}^{3/4}  \log(n\vee p) \{(\log p)/n\}^{1/4}  = o(1)$.
\end{enumerate}
Condition \ref{condition:error_bound_quantile} includes the estimation error bound of $\wh\btheta_{Q,k}$ derived in the previous literature \citep{belloni2011l1}. A similar condition is imposed in \cite{zhao2014general} for testing high dimensional composite quantile regression coefficients. 
When $ s_{X}$ and $s_{Q,Y}$ are finite, Condition \ref{condition:order_quantile} is equivalent to $\log p = o(n^{1/5})$. This condition has been imposed in \citet{belloni2014uniform} and \citet{zhao2014general}, which focus on the inference of the high-dimensional quantile regression model when $\epsilon$ is sub-Gaussian. 

Denote $\sigma^{2}_{Q,k}\defby (\tau - \tau^2) \E\{(X_{k} - \bgamma_{k}\trans\z_{k} )^2\}$.
{\theo{\label{theorem:quantile_normality}} Suppose Conditions \ref{condition:sub_gaussion}, \ref{condition:error_distribution}, \ref{condition:error_bound_quantile},
	and \ref{condition:order_quantile} are valid.
	\begin{enumerate}
		\item[(i)] 	 Under $H_{0,k}$, $n^{1/2}\wh\pQcov_{\tau}(Y, X_{k}\mid \z_{k})\stackrel{d}{\longrightarrow}{\cal N}(0, \sigma_{Q,k}^2) $ as $n\to\infty$.
		\item[(ii)] Under $H_{1,k}$, $\wh\pQcov_{\tau}(Y, X_{k}\mid \z_{k}) \stackrel{}{\longrightarrow} \pQcov_{\tau}(Y, 	X_{k}\mid \z_{k})$  as $n \to \infty$.
\end{enumerate}}
To conduct hypothesis test, we determine whether to reject $H_{0,k}$ in \eqref{test} based on the value of $n^{1/2}\wh\pQcov_{\tau}(Y, X_{k}\mid \z_{k}) /\wh\sigma_{Q,k}$, where  $\wh\sigma_{Q,k}^2\defby (\tau - \tau^2)\big\{n^{-1}\sum_{i=1}^{n}(X_{ki}-\wh\bgamma_{k}\trans\z_{ki})^2\big\}$.  
This test, using \( n^{1/2}\wh\pQcov_{\tau}(Y, X_{k}\mid \z_{k})/\wh\sigma_{Q,k} \) for \eqref{test}, is more restrictive than the proposed test based on partial Gini covariance, which requires \( \log(p) = o(n^{1/3}) \), as it necessitates \( \log(p) = o(n^{1/5}) \).


An issue arises when assessing the effectiveness of $\wh\pQcov_{\tau}(Y, X_{k}\mid \z_{k})$. Denote $\mu_{Q}(\beta_{k}^{\ast})\defby \pQcov_{\tau}(Y, X_{k}\mid \z_{k})$. $\mu_{Q}^{\prime}(\beta_{k}^{\ast})$ goes to $0$ at $\beta_{k}^{\ast} = 0$. 
Consider a specific scenario where both  $X_{k}$ and $\varepsilon$ follow the standard normal distributions, and $X_{k}$ is independent with $\x_{-k}$.
 Rather than calculating $\mu_{Q}^{\prime}(0)$, we investigate the limiting behavior of $\mu_{Q}^{\prime}(n^{-1/2}\beta_{k,0})$ as $n\to\infty$. It is straightforward to derive
\beqr\label{equation:are_quantile}
\lim_{n\to\infty}\mu_{Q}^{\prime}(n^{-1/2}\beta_{k,0})\rightarrow 0.
\eeqr
This limit indicates that $\wh\pQcov_{\tau}(Y, X_{k}\mid \z_{k})$ might be too small under local alternatives when covariates are mutually independent, leading to poor performance in testing \eqref{test}. 
This aspect has not been previously addressed in the literature. We illustrate the empirical power for $\wh\pQcov_{\tau}(Y, X_{k}\mid \z_{k})$ in Section \ref{subsec:4.1}, which indeed confirms this observation. The derivation of Equation \eqref{equation:are_quantile} is provided in the Supplementary Material.




\section{Simulation Studies}\label{sec:4}
\subsection{Partial Gini Covariance Based Tests}\label{subsec:4.1}
In this section, we present the finite sample performance of the partial Gini covariance-based test.
We consider two studies for our analysis.
The first study aims to examine the empirical size of the partial Gini covariance-based test and partial quantile covariance-based test.
The second study focuses on comparing the power performance of other high-dimensional independence tests against others.
Throughout the simulations, we generate $\x_{i} = (X_{i1},\ldots, X_{ip})\trans$ for $i=1,\ldots,n$ independently from multivariate normal distribution with mean zero and covariance $\bSig$, where $(\bSig)_{s,t} = 0.5^{\abs{s-t}}$. We consider the following linear model, for $i=1,\ldots,n$,
\beqrs
Y_{i} = \x_{i}\trans\bbeta^{\ast} + \varepsilon_{i}.
\eeqrs
Here, $\bbeta^{\ast}$ here is set as $(\beta_{1}^{\ast},\ldots,\beta_{10}^{\ast}, \vz_{p-10})\trans$, and $\varepsilon_{i}, i=1,\ldots,n$ are generate independently from $\x_{i}$. We consider three different noise settings: 
(i) $\varepsilon\sim \calN(0,1)$, 
(ii) $\varepsilon\sim \textrm{T}_{2}(0,1)$, 
(iii) $\varepsilon\sim \textrm{Cauchy}(0,1)$.  $\textrm{T}_{2}(0,1)$ represents standard $t$ distribution with $2$ degrees of freedom.
The latter two are both heavy-tailed distributions.
We fix $n=200$ and consider $p=500$ and $p=2000$ under three different distributions of $\varepsilon_{i}$ respectively.

We compare the empirical distributions of six test statistics in the following context, which can be divided into two types.
The first type is partial covariance-based tests, involving partial Gini covariance-based test statistic, 
quantile partial covariance-based test statistic   with $\tau = 0.5$, and
partial Pearson covariance-based test statistic.
To ensure that linear coefficients $\btheta_{P,k}$ are correctly specified in partial Pearson covariance under heavy-tailed cases, we also implement another modified estimate by applying rank Lasso \citep{wang2020tuning}. The estimate is denoted as 
\beqrs
\wt\pPcov(Y, X_{k}\mid \z_{k}) = n^{-1}\sum_{i=1}^{n}(Y_{i}-\wh\btheta_{G,k}\trans\z_{ki})(X_{ki}-\wh\bgamma_{k}\trans\z_{ki}),
\eeqrs
The corresponding test statistic is $n^{1/2}\wt\pPcov(Y, X_{k}\mid \z_{k})/\wt\sigma_{P,k}$, where $\wt\sigma_{P,k}^2\defby \big\{n^{-1}\sum_{i=1}^{n}(Y_{i}-\wh\btheta_{G,k}\trans\z_{ki})^2\big\}\big\{n^{-1}\sum_{i=1}^{n}(X_{ki}-\wh\bgamma_{k}\trans\z_{ki})^2\big\}$.
The second group includes debiased Lasso \citep{zhang2014confidence} and its bootstrap version \cite{dezeure2017high}.
These two test statistics are implemented using R package \texttt{hdi} with $101$ times of bootstrap.

\noindent\textbf{Study 1.} In this study, we examine the empirical sizes of the aforementioned test statistics.
We set $\beta_{1}^{\ast}=\ldots=\beta_{10}^{\ast} = \beta$ and  vary $\beta$ in $\{0.2, 0.4, \ldots,1\}$.
We test the hypothesis $H_{0,11}$, which represents the independent case. Here, the significant level $\alpha$ is set to $0.05$.
The empirical sizes under six scenarios are presented in Table \ref{table:size}.

\begin{table}[htpb!]
	\captionsetup{font = footnotesize}
	\caption{The empirical sizes of  six test statistics when $\beta$ vary in $\{0.2, 0.4, \ldots, 1\}$: partial Gini covariance based tests ($\pGcov$); partial quantile covariance based tests ($\pQcov$); 
		partial Pearson covariance based tests ($\pPcov$);
		modified partial Pearson covariance based tests ($\pPcov_{m}$);
		de-biased Lasso based tests ($\dBeta$);
		de-biased Lasso based tests using bootstrap ($\dBeta_{b}$).
	}
	\label{table:size}    
	\centering
	\scalebox{0.9}{
		\begin{tabular}{llllll|lllll}
			\hline
			\hline
			$\beta$ & 0.2   & 0.4   & 0.6   & 0.8   & 1     & 0.2    & 0.4   & 0.6   & 0.8   & 1     \\
			\hline
			& \multicolumn{5}{c}{$(n,p) = (200, 500)$, $\varepsilon\sim \calN(0,1)$} & \multicolumn{5}{c}{$(n,p) = (200, 2000)$, $\varepsilon\sim \calN(0,1)$} \\
			\hline
			$\pGcov$ & 0.040  & 0.066 & 0.058 & 0.054 & 0.040  & 0.038   & 0.052 & 0.060  & 0.042 & 0.060 \\
			$\pQcov$ & 0.054 & 0.052 & 0.064 & 0.048 & 0.044 & 0.034  & 0.040 & 0.036  & 0.060  & 0.050 \\
			$\pPcov$ & 0.048 & 0.050  & 0.048 & 0.046 & 0.048 & 0.050  & 0.060  & 0.056 & 0.076 & 0.066 \\
			$\pPcov_{m}$ & 0.034 & 0.054 & 0.048 & 0.062 & 0.048 & 0.038  & 0.038 & 0.054 & 0.042 & 0.050  \\
			$\dBeta$ & 0.050 & 0.066 & 0.054 & 0.058 & 0.046 & 0.056  & 0.066 & 0.064 & 0.076 & 0.072  \\
			$\dBeta_{b}$ & 0.040 & 0.050 & 0.074 & 0.066 & 0.048 & 0.042  & 0.044 & 0.066 & 0.046 & 0.054  \\
			\hline
			& \multicolumn{5}{c}{$(n,p) = (200, 500)$, $\varepsilon\sim \textrm{T}_{2}(0,1)$}      & \multicolumn{5}{c}{$(n,p) = (200, 2000)$, $\varepsilon\sim \textrm{T}_{2}(0,1)$}      \\
			\hline
			$\pGcov$ & 0.050  & 0.060  & 0.062 & 0.054 & 0.052 & 0.052  & 0.062 & 0.056 & 0.062 & 0.060  \\
			$\pQcov$ & 0.050  & 0.054 & 0.060  & 0.038 & 0.060  & 0.054  & 0.054 & 0.056 & 0.060  & 0.056 \\
			$\pPcov$ & 0.050  & 0.044 & 0.054 & 0.044 & 0.042 & 0.038  & 0.076 & 0.048 & 0.068 & 0.066 \\
			$\pPcov_{m}$ & 0.026 & 0.032 & 0.030  & 0.030  & 0.042 & 0.026  & 0.042 & 0.032 & 0.040  & 0.028 \\
			$\dBeta$ & 0.048 & 0.056 & 0.066 & 0.058 & 0.076 & 0.052  & 0.082 & 0.066 & 0.092 & 0.100  \\
			$\dBeta_{b}$ & 0.066 & 0.078 & 0.070 & 0.046 & 0.054 & 0.058  & 0.074 & 0.062 & 0.076 & 0.064  \\
			\hline
			& \multicolumn{5}{c}{$(n,p) = (200, 500)$, $\varepsilon\sim \textrm{Cauchy}(0,1)$} & \multicolumn{5}{c}{$(n,p) = (200, 2000)$, $\varepsilon\sim \textrm{Cauchy}(0,1)$} \\
			\hline
			$\pGcov$ & 0.034 & 0.052 & 0.056 & 0.050  & 0.068 & 0.050   & 0.054 & 0.068 & 0.054 & 0.066 \\
			$\pQcov$ & 0.050  & 0.060  & 0.052 & 0.048 & 0.040  & 0.044  & 0.040  & 0.048 & 0.060  & 0.058 \\
			$\pPcov$ & 0.016 & 0.018 & 0.030  & 0.036 & 0.050  & 0.018  & 0.044 & 0.030  & 0.032 & 0.044 \\
			$\pPcov_{m}$ & 0.018 & 0.026 & 0.022 & 0.026 & 0.030  & 0.018  & 0.024 & 0.012 & 0.020  & 0.022 \\			
			$\dBeta$ & 0.034 & 0.070 & 0.058 & 0.064 & 0.060 & 0.074  & 0.058 & 0.050 & 0.052 & 0.068  \\
			$\dBeta_{b}$ & 0.042 & 0.070 & 0.070 & 0.074 & 0.068 & 0.088  & 0.062 & 0.056 & 0.062 & 0.072  \\
			\hline
			\hline
	\end{tabular}}
\end{table}

From the table, it is evident that the sizes of the proposed test and the test based on partial quantile covariance are well-controlled at approximately $0.05$ under three different noise settings.
In contrast, both of the two partial Pearson covariance-based test statistics are overly conservative when  $\varepsilon$ follows heavy-tailed distributions. 
In fact, partial Pearson covariance-based test statistics failed to converge to the normal distribution, because the asymptotic variances of estimated partial Pearson covariance $n^{1/2}\wh\pPcov(Y, X_{k}\mid \z_{k})$ and $n^{1/2}\wt\pPcov(Y, X_{k}\mid \z_{k})$, do not exist. 
As for two debiased Lasso-based test statistics,  the empirical sizes cannot be well-controlled when  $\varepsilon$ follows heavy-tailed distributions.
For instance, when $(n,p)=(200, 2000)$ and $\varepsilon$ follow $t$ distribution with $2$ degrees of freedom, the empirical size of debiased Lasso-based tests reach $0.100$.
This observation highlights the importance of employing rank-based partial covariance to achieve the asymptotic normality of the test.

\begin{figure}[htbp!]
	\captionsetup{font = footnotesize}
	\centering
	\subfigure[$(n,p) = (200, 500)$, $\varepsilon\sim \calN(0,1)$]{
		\begin{minipage}[t]{0.45\linewidth}
			\centering
			\includegraphics[width=2.6in]{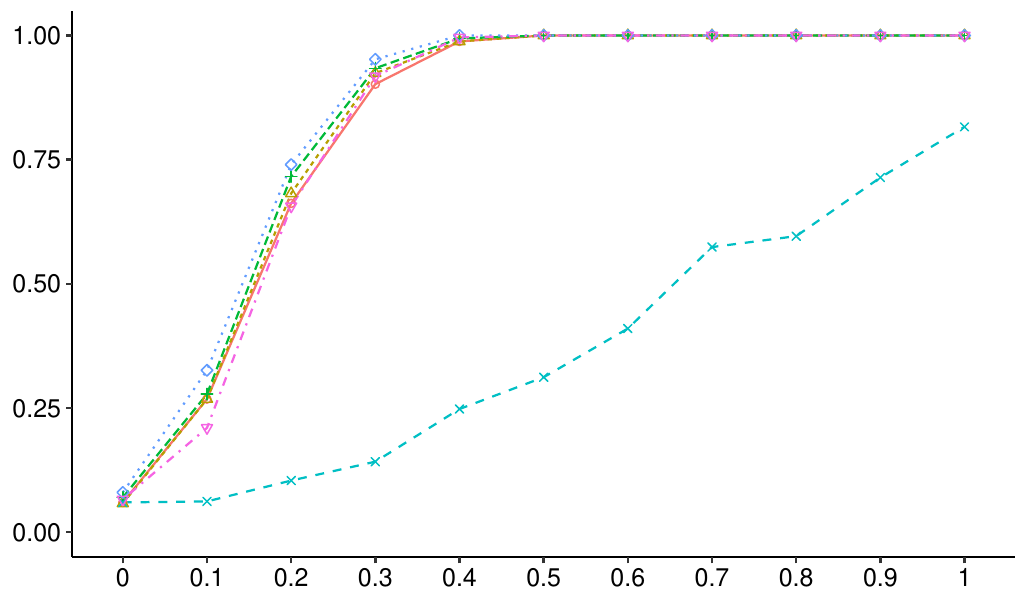}
		\end{minipage}%
	}%
	\subfigure[$(n,p) = (200, 2000)$, $\varepsilon\sim \calN(0,1)$]{
		\begin{minipage}[t]{0.45\linewidth}
			\centering
			\includegraphics[width=2.6in]{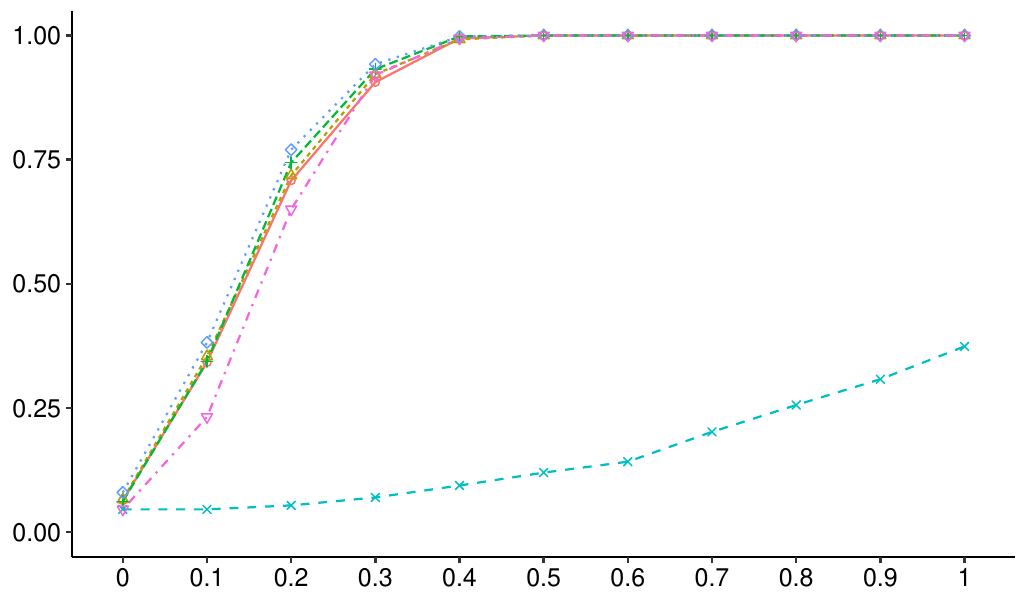}
		\end{minipage}%
	}%
	
	\subfigure[$(n,p) = (200, 500)$, $\varepsilon\sim t_{2}$]{
		\begin{minipage}[t]{0.45\linewidth}
			\centering
			\includegraphics[width=2.6in]{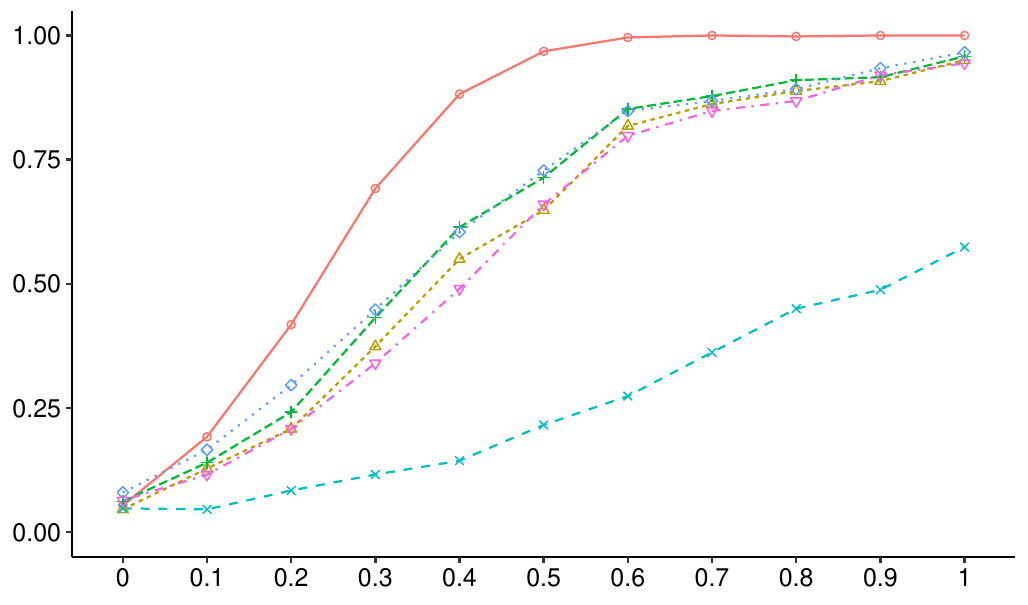}
		\end{minipage}%
	}%
	\subfigure[$(n,p) = (200, 2000)$, $\varepsilon\sim t_{2}$]{
		\begin{minipage}[t]{0.45\linewidth}
			\centering
			\includegraphics[width=2.6in]{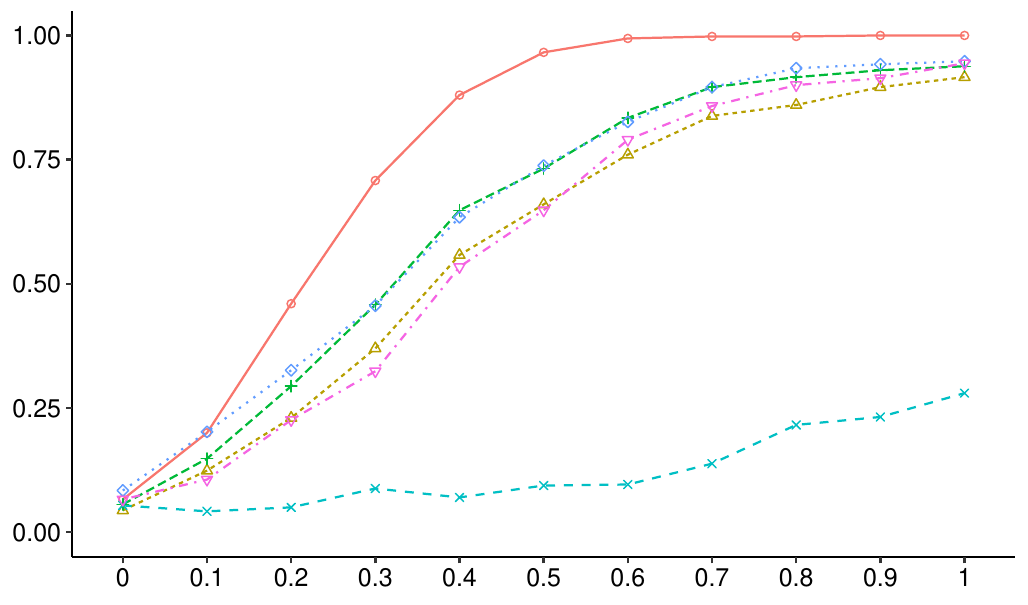}
		\end{minipage}%
	}%
	
	\subfigure[$(n,p) = (200, 500)$, $\varepsilon\sim \textrm{Cauchy}(0,1)$]{
		\begin{minipage}[t]{0.45\linewidth}
			\centering
			\includegraphics[width=2.6in]{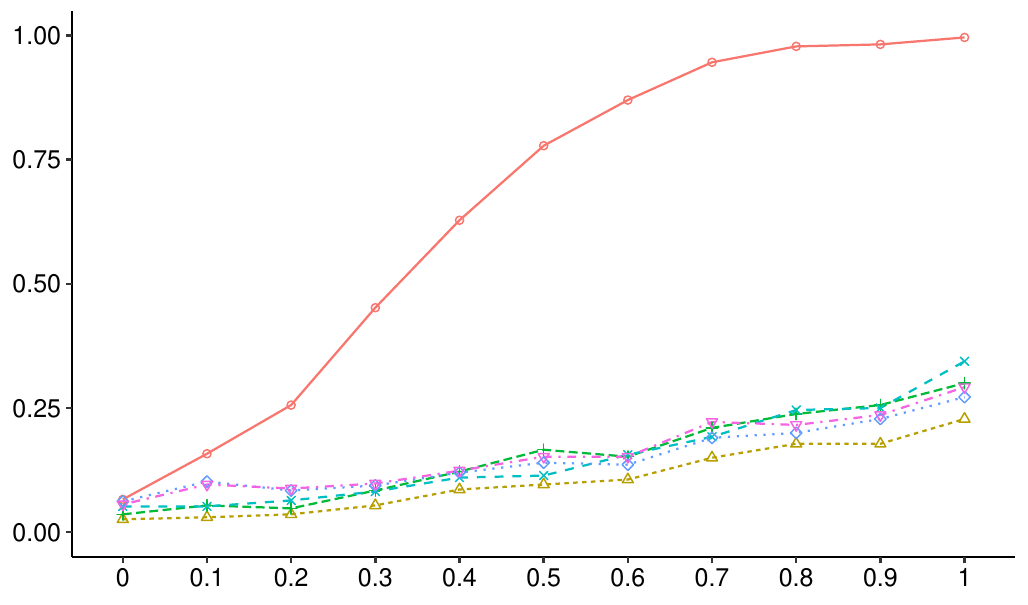}
		\end{minipage}%
	}%
	\subfigure[$(n,p) = (200, 2000)$, $\varepsilon\sim \textrm{Cauchy}(0,1)$]{
		\begin{minipage}[t]{0.45\linewidth}
			\centering
			\includegraphics[width=2.6in]{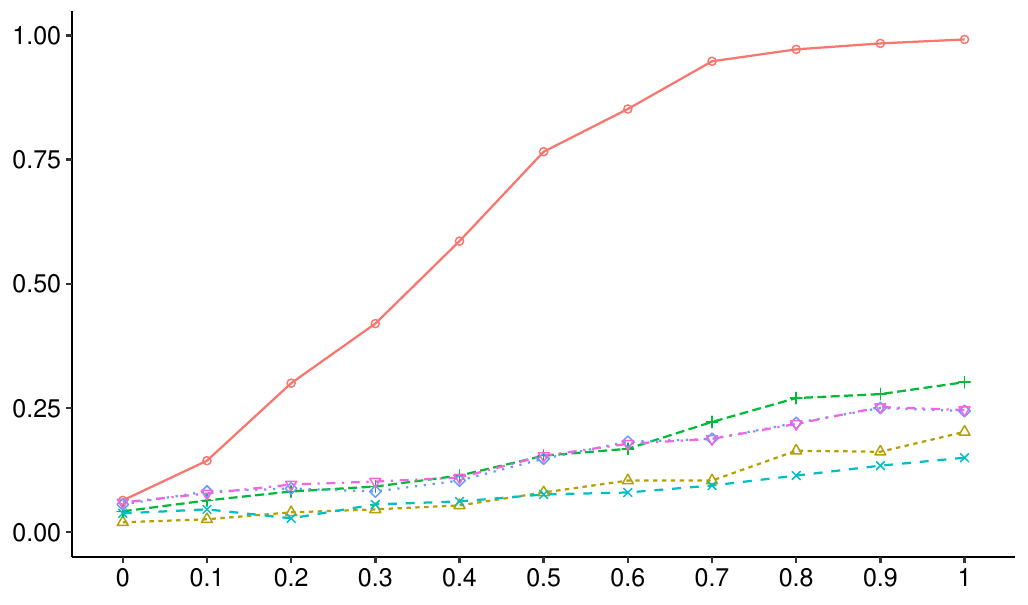}
		\end{minipage}%
	}%
	
	\subfigure{
		\begin{minipage}[t]{0.85\linewidth}
			\centering
			\includegraphics[width=2.6in]{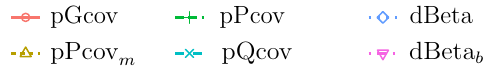}
		\end{minipage}
	}%
	\centering
	\caption{ The empirical powers of  six test statistics: partial Gini covariance based tests ($\pGcov$); partial quantile covariance based tests ($\pQcov$); 
		partial Pearson covariance based tests ($\pPcov$);
		modified partial Pearson covariance based tests ($\pPcov_{m}$);
		de-biased Lasso based tests ($\dBeta$);
		de-biased Lasso based tests using bootstrap ($\dBeta_{b}$).
		The horizontal axis represents the signal strength $\beta_{1}^{\ast}$.
	}
	\label{fig:power}
\end{figure}

\noindent\textbf{Study 2.} 
We examine the empirical power performance of the test statistics in Study 1. 
In this study, we test $H_{0,1}$.
We keep $\beta_{2}^{\ast}=\ldots=\beta_{10}^{\ast}=1$ and vary $\beta_{1}^{\ast}$ in $\{0, 0.1,\ldots,1\}$ to control the signal strength. 
When $\beta_{1}^{\ast}=0$, the null hypothesis is true. 
Here, we reject the null hypothesis at significance level $\alpha=0.05$.
The other settings remain the same as in Study 1.
The empirical powers are depicted in Figure \ref{fig:power}.  

From Figure \ref{fig:power}, 
we observe that as the signal strength $\beta_{1}^{\ast}$ increases, all the tests demonstrate improved power performance.
Notably, when the noise follows the standard normal distribution, the partial Gini covariance-based test shows power performance comparable to the partial Pearson-based tests and two debiased Lasso-based tests. 
The power of the test based on partial quantile covariance grows slowly as $\beta^{\ast}_{1}$ increases, compared with the other five methods. This phenomenon can be attributed to equation \eqref{equation:are_quantile}, which quantifies Pitman's efficiency of partial quantile covariance.
When noise follows $t$ or Cauchy distribution, the power of our proposed partial Gini covariance outperforms the other five methods in terms of power.  
The simulation results echo the discussion about asymptotic relative efficiency in equation \eqref{equation:are}.
In addition, Figure \ref{fig:power} illustrates that the two partial Pearson covariance tests and two de-biased Lasso tests display similar empirical power performance.

\subsection{Partial Gini Chi-square Tests}\label{subsec:4.2}
We investigate the empirical sizes and powers of chi-square test statistics based on multivariate partial Gini covariance, with the same high-dimensional model as discussed in the previous subsection.
In this study, we test the hypothesis $H_{0,\calS}$.
To be specific, we consider two cases:
\begin{enumerate}
	\item[(1)]  $\calS=\{1,2,3\}$, $\beta_{4}^{\ast}=\ldots=\beta_{10}^{\ast}=1$;
	\item[(2)]  $\calS=\{1,\ldots,10\}$, $\beta_{4}^{\ast}=\ldots=\beta_{10}^{\ast}=0$.
\end{enumerate}
In the first case, we test $3$ elements, with all elements greater than $0$ under the alternative hypothesis.
The second case represents a sparse scenario, where only 3 out of the 10 elements are non-zero.
We vary $\beta=\beta_{1}^{\ast}= \beta_{2}^{\ast}= \beta_{3}^{\ast}$ in $\{0, 0.1,\ldots,1\}$ to control the dependence level. Specifically, when $\beta=0$, the null hypothesis holds true. 
Here, we reject the null hypothesis at significance level $\alpha=0.05$.
The empirical sizes and powers are reported in Table \ref{table:multi_sizepower}.
\begin{table}[htpb!]
	\captionsetup{font = footnotesize}
	\caption{The empirical sizes powers of our proposed chi-square tests based on multivariate partial Gini covariance.
		Three different distributions of $\varepsilon$ are considered: standard normal distribution ($\calN(0,1)$); standard $t$ distribution with $2$ degrees of freedom ($\textrm{T}_{2}(0,1)$); standard Cauchy distribution ($\textrm{Cauchy}(0,1)$).
	}
	\label{table:multi_sizepower}    
	\centering
	\scalebox{0.9}{
		\begin{tabular}{llllllllllll}
			\hline
			\hline
			$\beta$	& 0     & 0.1   & 0.2   & 0.3   & 0.4   & 0.5   & 0.6   & 0.7   & 0.8   & 0.9   & 1     \\
			\hline
			\multicolumn{12}{c}{Case 1: $\calS = \{1,2,3\}$, $\beta_{4}^{\ast}=\ldots=\beta_{10}^{\ast}= 1$} \\
			\hline
			\multicolumn{12}{c}{$(n,p)=(200, 500)$}       \\
			\hline                                               
			$\calN(0,1)$ & 0.056 & 0.492 & 0.972 & 1.000      & 1.000      & 1.000      & 1.000      & 1.000      & 1.000      & 1.000      & 1.000      \\
			$\textrm{T}_2(0,1)$ & 0.050  & 0.224 & 0.742 & 0.970  & 0.998 & 1.000      & 1.000      & 1.000      & 1.000      & 1.000      & 1.000      \\
			$\textrm{Cauchy}(0,1)$ & 0.052 & 0.158 & 0.402 & 0.700   & 0.906 & 0.968 & 0.996 & 1.000      & 1.000      & 1.000      & 1.000    \\
			\hline
			\multicolumn{12}{c}{$(n,p)=(200, 2000)$}                                                     \\
			\hline
			$\calN(0,1)$ & 0.050  & 0.200   & 0.728 & 0.972 & 1.000      & 1.000      & 1.000      & 1.000      & 1.000      & 1.000      & 1.000      \\
			$\textrm{T}_2(0,1)$ & 0.062 & 0.122 & 0.404 & 0.794 & 0.950  & 0.992 & 0.998 & 1.000      & 1.000      & 1.000      & 1.000      \\
			$\textrm{Cauchy}(0,1)$ & 0.060  & 0.096 & 0.190  & 0.404 & 0.624 & 0.816 & 0.926 & 0.952 & 0.976 & 0.984 & 0.998 \\
			\hline
			\multicolumn{12}{c}{Case 2: $\calS = \{1,\ldots,10\}$, $\beta_{4}^{\ast}=\ldots=\beta_{10}^{\ast}= 0$} \\
			\hline
			\multicolumn{12}{c}{$(n,p)=(200, 500)$}       \\
			\hline                                               
			$\calN(0,1)$ & 0.058 & 0.392 & 0.956 & 1.000      & 1.000      & 1.000      & 1.000      & 1.000      & 1.000      & 1.000      & 1.000      \\
			$\textrm{T}_2(0,1)$ & 0.040  & 0.150 & 0.668 & 0.936  & 0.998 & 1.000      & 1.000      & 1.000      & 1.000      & 1.000      & 1.000      \\
			$\textrm{Cauchy}(0,1)$ & 0.052 & 0.088 & 0.270 & 0.624   & 0.862 & 0.936 & 0.994 & 0.998      & 0.998      & 1.000      & 1.000    \\
			\hline
			\multicolumn{12}{c}{$(n,p)=(200, 2000)$}                                                     \\
			\hline
			$\calN(0,1)$ & 0.046  & 0.284   & 0.850 & 0.992 & 1.000      & 1.000      & 1.000      & 1.000      & 1.000      & 1.000      & 1.000      \\
			$\textrm{T}_2(0,1)$ & 0.036 & 0.142 & 0.446 & 0.820 & 0.970  & 0.986 & 1.000 & 1.000      & 1.000      & 1.000      & 1.000      \\
			$\textrm{Cauchy}(0,1)$ & 0.052  & 0.064 & 0.144  & 0.346 & 0.578 & 0.776 & 0.872 & 0.936 & 0.976 & 0.988 & 0.994 \\
			\hline
			\hline
	\end{tabular}}
\end{table}
Table \ref{table:multi_sizepower} confirms that the sizes of proposed methods are well-controlled in both cases, given the fact that all the empirical sizes are around $0.05$ when $\beta=0$. When $\beta>0$, we can find that the empirical power has better performance as the dependence strength level $\beta$ increases. The empirical powers of our tests approach $1$ as $\beta$ increases.
\section{Real Data Application}\label{sec:5}
In this section, we apply our methods to a car pricing dataset.
Our goal is to identify the variables that are significant in predicting the car price.
The dataset comes from market surveys conducted across the United States by the Chinese automobile company Geely Auto and  is available for download from
\url{https://www.kaggle.com/datasets/goyalshalini93/car-data}.
The data comprises $n=205$ different cars with $25$ variables including the price and other $24$  variables related to the car characteristic.
We remove $5$ description variables such as car names and focus on the remaining $19$ predictors. 

We consider six methods, as discussed in Section \ref{subsec:4.1}, which include partial Gini covariance, quantile partial covariance with $\tau = 0.5$, partial Pearson covariance, and partial Pearson covariance modified-based test statistics.
To transform the real dataset into a high-dimensional one and further examine our methods in controlling type I error, we artificially create $380$ variables obtained by $20$ different permutations of the row order of the original $19$ predictors.
This results in a dimensionality of $p = 380+19 = 399$.
For each true covariate, we calculate the partial covariances-based test statistics and the corresponding $p$-values.
To examine whether our methods can control the empirical size, we set different significance levels $\alpha\in\{0.01, 0.05, 0.1\}$ and report the rejection proportions among $380$ simulated variables in Table \ref{table:realdata_size}.
\begin{table}[htpb!]
	\captionsetup{font = footnotesize}
	\caption{
		The rejection proportions among $380$ simulated variables at significance levels $\alpha\in\{0.01, 0.5, 0.1\}$.
		Six independence test statistics include: partial Gini covariance ($\pGcov$),
		partial quantile  covariance ($\pQcov$) with $\tau=0.5$,
		partial Pearson covariance ($\pPcov$), 
		modified partial Pearson covariance ($\pPcov_{m}$),
		de-biased Lasso ($\dBeta$), and
		de-biased Lasso using bootstrap based tests ($\dBeta_{b}$).
	}
	\label{table:realdata_size}    
	\centering
	\scalebox{1}{
	\begin{tabular}{lrrr|lrrr}
		\hline
		\hline
		$\alpha$	& 0.01  & 0.05  & 0.1 && 0.01  & 0.05  & 0.1    \\
		\hline
		$\pGcov$ & 0.011 & 0.053 & 0.105  & $\pQcov$ & 0.008 & 0.061 & 0.113 \\
		$\pPcov$ & 0.003 & 0.024 & 0.061 & $\pPcov_{m}$ & 0.008 & 0.042 & 0.089  \\
		$\dBeta$ & 0.008 & 0.037 & 0.076 & $\dBeta_{b}$ & 0.011 & 0.031 & 0.076  \\
		\hline
		\hline
	\end{tabular}}
\end{table}
From the table, we can see that the rejection proportions of partial Gini covariance and partial quantile covariance are around the significance levels, which indicates that these two methods work well in controlling type I errors.
In contrast,  tests based on partial Pearson covariance and debiased Lasso tend to be conservative. This is mainly due to the heavy-tailed distribution of the response variable.

To further identify the variables that are significant in predicting the price, we report the test statistics and corresponding $p$-values for $19$ original variables in Table \ref{table:realdata_pvalue}.
\begin{table}[htpb!]
	\captionsetup{font = footnotesize}
	\caption{The test statistics and corresponding $p$-values for $19$ original variables.
		Six independence test statistics include partial Gini covariance ($\pGcov$),
		partial quantile  covariance ($\pQcov$) with $\tau=0.5$,
		partial Pearson covariance ($\pPcov$), 
		modified partial Pearson covariance ($\pPcov_{m}$),
		de-biased Lasso ($\dBeta$), and
		de-biased Lasso using bootstrap based tests ($\dBeta_{b}$).
	}
	\label{table:realdata_pvalue}    
	\centering
	\scalebox{0.75}{
		\begin{tabular}{lrrrrr}
			\hline
			\hline    
			&  symboling      & fueltype   & aspiration & doornumber & enginelocation   \\
			$\pGcov$ & 1.841 (0.066)     & -1.952 (0.051)    & -0.647 (0.517)    & -0.295 (0.768)    & -2.086  (0.037)         \\
			$\pQcov$ & 1.209 (0.226)     & -0.823 (0.411)    & 0.025 (0.979)    & 0.334 (0.738)    & -1.992  (0.046)         \\
			$\pPcov$ & 0.554 (0.580)     & -0.753 (0.451)    & 1.140 (0.254)    & -0.593 (0.553)    & -1.840  (0.066)         \\
			$\pPcov_{m}$ & 1.009 (0.313)  & -1.327 (0.185)   & 0.859 (0.390)    & -1.142 (0.254)    & -1.516  (0.129)         \\
			$\dBeta$ & 0.994 (0.320)  & -1.129 (0.259)   & 0.987 (0.323)    & -0.656 (0.512)    & -6.431  ($<$0.001)         \\
			$\dBeta_{b}$ & 0.994 (0.598)  & -1.129 (0.500)   & 0.987 (0.147)    & -0.656 (0.618)    & -6.431  (0.010)         \\	    
			\hline
			& wheelbase      & carlength  & carwidth   & carheight  & curbweight       \\
			$\pGcov$ & 1.057 (0.290)    & 2.181 (0.029)    & 2.543 (0.011)      & -0.222 (0.824)    & 5.069 ($<$0.001)           \\
			$\pQcov$ & 1.735 (0.083)    & 1.381 (0.167)    & 3.019 (0.003)      & 0.924 (0.356)    & 4.086 ($<$0.001)           \\
			$\pPcov$ & 1.778 (0.075)    & -1.455 (0.146)    & 2.713 (0.007)      & 0.488 (0.626)    & 1.739 (0.082)           \\
			$\pPcov_{m}$ & 0.918 (0.359)  & 1.116 (0.264)    & 1.894 (0.058)    & -0.669 (0.505)    & 2.745 (0.006)           \\
			$\dBeta$ & 1.734 (0.079)    & -0.701 (0.483)   & 4.784 ($<$0.001)       & 0.545 (0.586)    & 3.558  ($<$0.001)         \\
			$\dBeta_{b}$ & 1.734 (0.029)    & -0.701 (0.049)   & 4.784 (0.010)       & 0.545 (0.245)    & 3.558  (0.029)        \\	  
			\hline
			& cylindernumber & enginesize & boreratio  & stroke     & compressionratio \\
			$\pGcov$ & 2.415 (0.016)    & 1.757 (0.079)      & 1.474 (0.140)      & -0.193 (0.847)    & 0.177 (0.859)     \\
			$\pQcov$ & -1.191 (0.234)   & 0.966 (0.334)      & -0.731 (0.465)      & -1.547 (0.122)    & 1.684 (0.092)     \\
			$\pPcov$ & -1.784 (0.074)    & 2.555 (0.011)      & -2.580 (0.010)      & -3.026 (0.002)    & 0.122 (0.902)     \\
			$\pPcov_{m}$ & 2.024 (0.043)    & 1.742 (0.081)    & 0.811 (0.418)     & 0.124 (0.901)    & 0.007 (0.995)     \\
			$\dBeta$ & -1.478 (0.139)  & 8.024 ($<$0.001)   & -1.711 (0.087)    & -1.932 (0.053)    & -0.708  (0.479)         \\
			$\dBeta_{b}$ & -1.478 (0.108)  & 8.024 (0.010)   & -1.711 (0.010)    & -1.932 (0.010)    & -0.708  (0.735)          \\	  
			\hline
			& horsepower     & peakrpm    & citympg    & highwaympg &                  \\
			$\pGcov$  & 3.248 (0.001)          & 1.352 (0.176)      & -2.622 (0.009)     & -0.534 (0.594)     &                  \\
			$\pQcov$  & 3.376 (0.001)          & 0.937 (0.349)      & -1.340 (0.174)     & 0.242 (0.809)     &                  \\
			$\pPcov$  & 1.356 (0.175)         & 2.434 (0.015)      & -0.234 (0.815)     & -0.288 (0.773)     &                  \\
			$\pPcov_m$  & 1.408 (0.159)          & 1.165 (0.244)      & -0.753 (0.451)     & -1.029 (0.303)     &                  \\
			$\dBeta$ & 2.656 (0.008)       & 2.411 (0.016)   & -0.503 (0.615)    & -0.187 (0.851)    &         \\
			$\dBeta_{b}$ & 2.656 (0.068)       & 2.411 (0.010)   & -0.503 (0.657)    & -0.187 (0.794)    &            \\	
			\hline
			\hline            
	\end{tabular}}
\end{table}
From the results in Table \ref{table:realdata_pvalue}, we observe that our method has selected $7$ variables with $p$-values less than $0.05$.
Notably, these $7$ variables include the $4$ variables selected by partial quantile covariance and $2$ by modified partial Pearson covariance. 
This phenomenon indicates that our method has larger power in testing compared with other methods, which echoes the theoretical and simulation results in Section \ref{sec:4}.
Due to the heavy-tailed distribution of the response, the partial Pearson covariance-based test does not work well and has inconsistent results with the previously mentioned three methods. 
The two debiased Lasso methods, despite employing the same test statistics, produce distinct $p$-values. This inconsistency indicates the price follows heavy-tailed distribution and the debiased Lasso methods are not applicable.

We further conduct our partial Gini chi-square test to examine its performance when the coefficients are multivariate. Specifically, we consider three different sets: (1) $\calS_{1} = \{\text{``curbweight"}, \text{``horsepower"}, \text{``citympg"}\}$ includes variables whose $p$-value is less than $0.01$ in pratial Gini covariance test.; (2) $\calS_{2} = \{\text{``enginelocation"}, \text{``carlength"}, \text{``carwidth"}, \text{``curbweight"}, $\\$ \text{``cylindernumber"},
\text{``horsepower"}, \text{``citympg"}\}$ includes variables whose $p$-value is less than $0.05$ in pratial Gini covariance test.
(3) $\calS_{3}$ is all $19$ original variables. 
We report the test statistics and corresponding $p$-values of these three tests in Table \ref{table:realdata_multi}, with all three $p$-values less than $0.001$. This implies that among all three subsets of variables, at least one is significant in predicting the car price, which is consistent with the results shown above.

\begin{table}[htbp!]
	\captionsetup{font = footnotesize}
	\caption{ 
		The partial Gini covariance chi-square test statistics ($W_{G}$) and their corresponding $p$-values.
	}
	\label{table:realdata_multi}    
	\centering
	\scalebox{1}{
	\begin{tabular}{lrrr}
		\hline
		\hline
		& $\calS_{1}$  & $\calS_{2}$  & $\calS_{3}$    \\
		\hline
		$W_{G}$ & 42.530 & 57.515 & 163.813 \\
		$p$-values & $<$0.001 & $<$0.001 & $<$0.001 \\
		\hline
		\hline
	\end{tabular}}
\end{table}

\bigskip
\begin{center}
{\large\bf SUPPLEMENTARY MATERIAL}
\end{center}
The supplementary materials include all the technical proofs.
%
%
%
%

\bibliographystyle{agsm}
\bibliography{reference}
\end{document}